\documentclass[%
 reprint,
 superscriptaddress,
 amsmath,amssymb,
 aps,
]{revtex4-1}

\usepackage{graphicx}
\usepackage{dcolumn}
\usepackage{bm}

\begin{document}

\preprint{APS/123-QED}

\title{$\phi$ photoproduction on the proton at $E_{\gamma}$ = 1.5 - 2.9 GeV}

\author{K.~Mizutani}
\author{M.~Niiyama}
\affiliation{Department of Physics, Kyoto University, Kyoto 606-8502, Japan}
\author{T.~Nakano}
\author{M.~Yosoi}
\author{Y.~Nozawa}
\affiliation{Research Center for Nuclear Physics, Osaka University, Ibaraki, Osaka 567-0047, Japan}
\author{D.~S.~Ahn}
\affiliation{RIKEN, The Institute of Physical and Chemical Research, Wako, Saitama 351-0198, Japan}
\author{J.~K.~Ahn}
\affiliation{Department of Physics, Korea University, Seoul 02841, Republic of Korea}
\author{W.~C.~Chang}
\affiliation{Institute of Physics, Academia Sinica, Taipei 11529, Taiwan}
\author{J.~Y.~Chen}
\affiliation{Light Source Division, National Synchrotron Radiation Research Center, Hsinchu 30076, Taiwan}
\author{S.~Dat\'e}
\affiliation{Japan Synchrotron Radiation Research Institute, Sayo, Hyogo 679-5143, Japan}
\author{W.~Gohn}
\affiliation{Department of Physics, University of Connecticut, Storrs, Connecticut 06269-3046, USA}
\author{H.~Hamano}
\affiliation{Research Center for Nuclear Physics, Osaka University, Ibaraki, Osaka 567-0047, Japan}
\author{T.~Hashimoto}
\affiliation{Department of Physics, Kyoto University, Kyoto 606-8502, Japan}
\author{K.~Hicks}
\affiliation{Department of Physics and Astronomy, Ohio University, Athens, Ohio 45701, USA}
\author{T.~Hiraiwa}
\author{T.~Hotta}
\affiliation{Research Center for Nuclear Physics, Osaka University, Ibaraki, Osaka 567-0047, Japan}
\author{S.~H.~Hwang}
\affiliation{Korea Research Institute of Standards and Science (KRISS), Daejeon 34113, Republic of Korea}
\author{T.~Ishikawa}
\affiliation{Research Center for Electron Photon Science, Tohoku University, Sendai, Miyagi 982-0826, Japan}
\author{K.~Joo}
\affiliation{Department of Physics, University of Connecticut, Storrs, Connecticut 06269-3046, USA}
\author{W.~S.~Jung}
\affiliation{Department of Physics, Korea University, Seoul 02841, Republic of Korea}
\author{Y.~Kato}
\affiliation{Kobayashi-Maskawa Institute, Nagoya University, Nagoya, Aichi 464-8602, Japan}
\author{H.~Katsuragawa}
\affiliation{Research Center for Nuclear Physics, Osaka University, Ibaraki, Osaka 567-0047, Japan}
\author{M.~H.~Kim}
\author{S.~H.~Kim}
\affiliation{Department of Physics, Korea University, Seoul 02841, Republic of Korea}
\author{H.~Kohri}
\author{Y.~Kon}
\affiliation{Research Center for Nuclear Physics, Osaka University, Ibaraki, Osaka 567-0047, Japan}
\author{H.~S.~Lee}
\affiliation{Department of Physics and Astronomy, Seoul National University, Seoul 151-742, Republic of Korea}
\author{Y.~Maeda}
\affiliation{Photon Therapy Center, Fukui Prefectural Hospital, Fukui 910-8526, Japan}
\author{Y.~Matsumura}
\affiliation{Research Center for Electron Photon Science, Tohoku University, Sendai, Miyagi 982-0826, Japan}
\author{T.~Mibe}
\affiliation{High Energy Accelerator Research Organization (KEK), Tsukuba, Ibaraki 305-0801, Japan}
\author{M.~Miyabe}
\affiliation{Research Center for Electron Photon Science, Tohoku University, Sendai, Miyagi 982-0826, Japan}
\author{Y.~Morino}
\affiliation{High Energy Accelerator Research Organization (KEK), Tsukuba, Ibaraki 305-0801, Japan}
\author{N.~Muramatsu}
\affiliation{Research Center for Electron Photon Science, Tohoku University, Sendai, Miyagi 982-0826, Japan}
\author{Y.~Nakatsugawa}
\affiliation{Institute of High Energy Physics, Chinese Academy of Sciences, Beijing 100049, China}
\author{H.~Noumi}
\affiliation{Research Center for Nuclear Physics, Osaka University, Ibaraki, Osaka 567-0047, Japan}
\author{Y.~Ohashi}
\affiliation{Japan Synchrotron Radiation Research Institute, Sayo, Hyogo 679-5143, Japan}
\author{T.~Ohta}
\affiliation{Department of Radiology, The University of Tokyo Hospital, Tokyo 113-8655, Japan}
\author{M.~Oka}
\affiliation{Research Center for Nuclear Physics, Osaka University, Ibaraki, Osaka 567-0047, Japan}
\author{J.~B.~Park}
\affiliation{Department of Physics, Korea University, Seoul 02841, Republic of Korea}
\author{J.~D.~Parker}
\affiliation{Neutron Science and Technology Center, Comprehensive Research Organization for Science and Society (CROSS), Tokai, Ibaraki 319-1106, Japan}
\author{C.~Rangacharyulu}
\affiliation{Department of Physics and Engineering Physics, University of Saskatchewan, Saskatoon, Saskatchewan S7N 5E2, Canada}
\author{S.~Y.~Ryu}
\author{Y.~Sada}
\affiliation{Research Center for Nuclear Physics, Osaka University, Ibaraki, Osaka 567-0047, Japan}
\author{T.~Sawada}
\affiliation{Physics Department, University of Michigan, Michigan 48109-1040, USA}
\author{T.~Shibukawa}
\affiliation{Department of Physics, The University of Tokyo, Tokyo 113-0033, Japan}
\author{S.~H.~Shiu}
\affiliation{Institute of Physics, Academia Sinica, Taipei 11529, Taiwan}
\author{Y.~Sugaya}
\affiliation{Research Center for Nuclear Physics, Osaka University, Ibaraki, Osaka 567-0047, Japan}
\author{M.~Sumihama}
\affiliation{Department of Education, Gifu University, Gifu 501-1193, Japan}
\author{S.~Tanaka}
\affiliation{Research Center for Nuclear Physics, Osaka University, Ibaraki, Osaka 567-0047, Japan}
\author{A.~O.~Tokiyasu}
\affiliation{Research Center for Electron Photon Science, Tohoku University, Sendai, Miyagi 982-0826, Japan}
\author{N.~Tomida}
\author{H.~N.~Tran}
\affiliation{Research Center for Nuclear Physics, Osaka University, Ibaraki, Osaka 567-0047, Japan}
\author{T.~Tsunemi}
\affiliation{Graduate School of Science, Kyoto University, Kyoto 606-8502, Japan}
\author{M.~Uchida}
\affiliation{Department of Physics, Tokyo Institute of Technology, Tokyo 152-8551, Japan}
\author{M.~Ungaro}
\affiliation{Department of Physics, University of Connecticut, Storrs, Connecticut 06269-3046, USA}
\author{Y.~Yanai}
\affiliation{Research Center for Nuclear Physics, Osaka University, Ibaraki, Osaka 567-0047, Japan}

\collaboration{LEPS Collaboration}

\date{\today}

\begin{abstract}
Differential cross sections at $t=t_{\text{min}}$ and decay asymmetries for the $\gamma p\rightarrow\phi p$ reaction have been measured using linearly polarized photons in the range 1.5 to 2.9 GeV.
These cross sections were used to determine the Pomeron strength factor.
The cross sections and decay asymmetries are consistently described by the $t$-channel Pomeron and pseudoscalar exchange model in the $E_{\gamma}$ region above 2.37 GeV.
In the lower energy region, an excess over the model prediction is observed in the energy dependence of the differential cross sections at $t=t_{\text{min}}$.
This observation suggests that additional processes or interference effects between Pomeron exchange and other processes appear near the threshold region.
\begin{description}
\item[PACS numbers]
13.60.Le, 25.20.Lj, 14.40.Be
\end{description}
\end{abstract}

\pacs{Valid PACS appear here}
\maketitle

Multi-gluon-exchange processes are universal for all the hadronic reactions, since the gluons are flavour-blind.
At low energies, meson-exchange processes are dominant, making it difficult to access the gluonic interactions in the $\rho$ and $\omega$ photoproductions.
Diffractive $\phi$-meson photoproduction is of particular interest in that the meson-exchange processes are suppressed due to the Okubo-Zweig-Iizuka rule, and can be a useful tool to study gluon dynamics or the Pomeron exchange process in the low-energy region.
Here, the term "Pomeron" expresses the Regge trajectory obtained with the model in Ref.~\cite{Titov2003}, where the spin dependence comes from two gluon exchange and the energy dependence comes from traditional Regge theory.

Phenomenologically, $\phi$-meson production cross sections at forward angles are characterized by the following diffractive exchange parameters $B$ and $(d\sigma/dt)_{t=t_{\text{min}}}$:
\begin{equation}
  \frac{d\sigma}{dt}=\left(\frac{d\sigma}{dt}\right)_{t=t_{\text{min}}}\exp[B(t-t_{\text{min}})],
\label{eq:exp}
\end{equation}
where $t_{\text{min}}$ denotes $t$ at zero degrees.
The differential cross sections at zero degrees $(d\sigma/dt)_{t=t_{\text{min}}}$ are predicted to increase monotonically with incident photon energy, because the dominant $t$-channel Pomeron exchange amplitude increases monotonically with the center of mass energy $\sqrt{s}$ \cite{Collins1977,Barone2002,Donnachie2002,Gribov2008}.
However, a nonmonotonic structure around $\sqrt{s}$ = 2.1 GeV was first reported by the LEPS Collaboration \cite{Mibe2005}, which cannot be explained by simple $t$-channel $\pi^0$, $\eta$ and Pomeron exchanges \cite{Titov2003}.
The CLAS Collaboration also confirmed this nonmonotonic structure by the extrapolation of their measurements at larger angles \cite{CLAS2014}.
They observed that energy dependence of the $t$-slope factor $B$ changes at around $\sqrt{s} = 2.3\ \text{GeV}$, and claimed that the production mechanism changes at around this energy.

Several production mechanisms have been suggested to explain the nonmonotonic structure, such as nucleon resonances \cite{Kiswandhi2010,Kiswandhi2012}, interference between $\phi p$ and $K^+\Lambda(1520)$, rescattering processes \cite{Ozaki2009,Ryu2014}, and additional gluonic processes \cite{Nakano1997,Kisslinger2000}.
Introducing nucleon resonances in the $s$-channel seems unlikely since the nonmonotonic structure observed by CLAS appears only at forward angles \cite{CLAS2014}.
A similar bump structure has been observed in the $\gamma p\rightarrow K^+\Lambda(1520)$ reaction \cite{Kohri2010}, which shares the same $K^+K^-p$ final state with the $\gamma p\rightarrow\phi p$ reaction.
This observation suggests that an interference effect could possibly explain the nonmonotonic structure of the $\phi$ photoproduction.
However, the LEPS measurement in 2016 \cite{Ryu2016} has shown that the interference effect is too small to account for the nonmonotonic structure.
The CLAS measurement of the neutral decay mode also supports the idea of a small interference effect \cite{CLASneut2014}.
As for the rescattering processes, Ryu \textit{et. al.} suggested that the nonmonotonic structure can be explained by taking into account the $K^+\Lambda(1520)$ rescattering process \cite{Ryu2014}.
However, they calculated only imaginary parts of the rescattering amplitudes, and introduced an artificial Pomeron-exchange suppression factor to enhance the rescattering effects near the threshold.
The possibility of the additional gluonic contributions near threshold has not been ruled out.

In LEPS 2005 measurement \cite{Mibe2005}, the maximum incident photon energy was 2.4 GeV, the same energy where CLAS claimed a change in the production mechanism.
In addition, the nonmonotonic structure measured by CLAS is stronger than what was observed in the LEPS measurement \cite{CLAS2014}.
To clarify this situation, we extended the energy range of the incident photon to 2.9 GeV \cite{Muramatsu2014}, and directly measured the forward scattered $\phi$ mesons using the LEPS dipole spectrometer.
Utilizing a linearly polarized photon beam, we also investigated spin observables that are sensitive to the spin parity of the exchanged particles in the $t$-channel \cite{Schilling1970}.
In this article, we present the cross sections at forward angles, the energy dependence of the $t$-slope factor $B$ and $(d\sigma/dt)_{t=t_{\text{min}}}$, and spin observables.

\begin{figure}[h]
\includegraphics[width=90mm,clip]{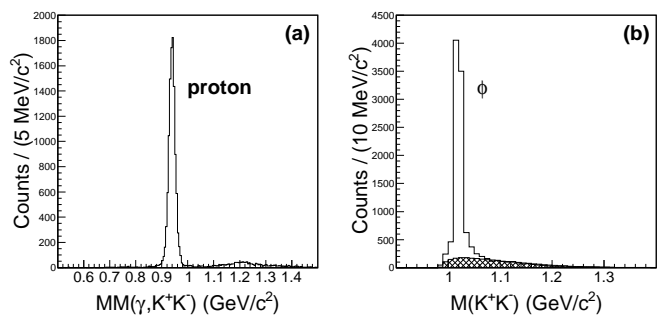}
\caption{\label{fig:phi_histo} (a) Missing mass distribution for the $p(\gamma,K^+K^-)X$ reaction. (b) The $K^+K^-$ invariant mass distribution after the proton selection cut on the $MM(\gamma,K^+K^-)$ distribution. The hatched histogram is the background distribution obtained by Monte Carlo simulations.}
\end{figure}

The experimental data were taken in 2007 and 2015 at SPring-8/LEPS in Japan \cite{Nakano2001}.
Linearly polarized photons with energies up to 2.9 GeV were produced by backward Compton scattering from the head-on collision between DUV laser photons and 8 GeV electrons in the storage ring.
The wavelengths of the DUV lasers are 257 nm and 266 nm for 2007 and 2015 data taking period, respectively.
The recoil electrons were detected in a tagging system near the collision point, giving the individual photon energies in the energy range from 1.5 to 2.9 GeV.
The energy resolution of the tagged photon was about 14 MeV.
The photon beam was incident on a 16-cm-long liquid hydrogen target.
The total number of photons on the target between 1.5 and 2.9 GeV was $8.0\times 10^{11}$.
The systematic uncertainties due to the number of photons and the target length were estimated to be 3\% and 1\%, respectively.
To detect $\phi$ mesons, $K^+$ and $K^-$ mesons produced at the target were momentum-analyzed by tracking devices and the dipole magnetic field.
The angular coverage of the LEPS spectrometer is about $\pm 0.4$ and $\pm 0.2$ rad in the horizontal and vertical directions, respectively.
Particle identification was made by reconstructing mass using time-of-flight and momentum information.
The $K^+K^-$ events were selected with a reconstructed mass spectrum within $4\sigma$ of the nominal mass value, where $\sigma$ is the momentum dependent mass resolution.
Since the most of the kaons from the $\phi$-meson decay have low momentum ($< 1.6$ GeV/$c$) due to the small $Q$-value (32.1 MeV), $\pi/K$ misidentification rate is small.
Reaction vertex points were reconstructed from the two kaon tracks, and used to select events in which the $\phi$ meson was produced at the target.
The missing mass distribution for the $p(\gamma,K^+K^-)X$ reaction ($MM(\gamma,K^+K^-)$) is shown in Fig.~\ref{fig:phi_histo}(a).
A clear peak corresponding to the proton is seen along with background events in which additional pions are produced.
The events with the $K^+K^-p$ in the final states were selected by requiring $0.85 < MM(\gamma,K^+K^-)<1.00\ \text{GeV}/c^2$.
Figure~\ref{fig:phi_histo}(b) shows the invariant mass distribution of $K^+K^-$ pairs for the events with the $K^+K^-p$ final states.
A peak corresponding to the $\phi$ meson is seen on top of the background.
We considered two sources of the background: nonresonant $K^+K^-p$ production and $\gamma p\rightarrow K^+\Lambda(1520)\rightarrow K^+K^-p$ reaction.
The background level was estimated by the simultaneous fit of the $K^+K^-$ invariant mass and $K^-p$ invariant mass distributions, using the mass distributions of $\phi p$, nonresonant $K^+K^-p$, and $K^+\Lambda(1520)$ reactions, which were obtained by Monte Carlo simulations with GEANT3 package \cite{Brun}.
The systematic errors due to the background estimation were $0.1-4.6\%$.
About 7000 $\gamma p\rightarrow\phi p$ events on the target were reconstructed.
The LEPS spectrometer acceptance including efficiencies for detectors and track reconstruction was calculated based on Monte Carlo simulations.

\begin{figure}[h]
\includegraphics[width=90mm,clip]{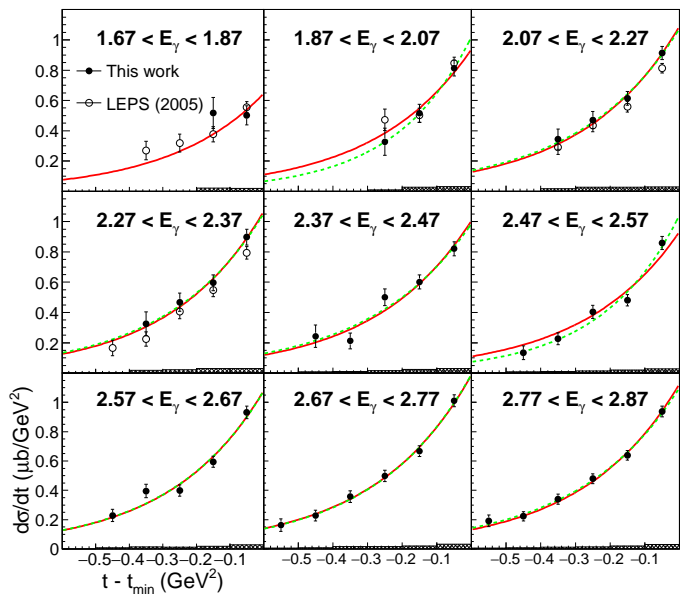}
\caption{\label{fig:cs_fit} The $t$ dependences of differential cross sections. The green dashed curves are the results of the fit using Eq.~(\ref{eq:exp}), with $(d\sigma/dt)_{t=t_{\text{min}}}$ and $B$ as floating parameters. The red solid curves are the results of the fit with fixing $B$ to $3.57\ \text{GeV}^{-2}$. The error bars represent statistical errors. The hatched histograms represent systematic errors.}
\end{figure}

\begin{figure}[h]
\includegraphics[width=90mm,clip]{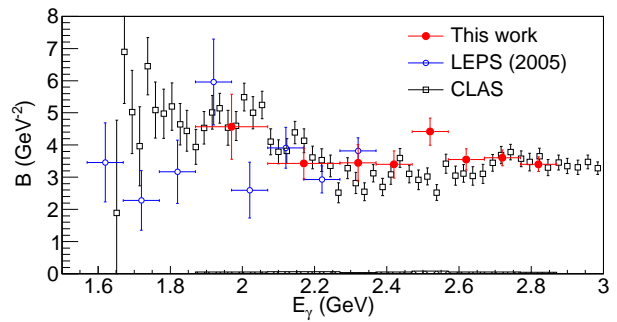}
\caption{\label{fig:bpar} Energy dependence of $t$-slope factor $B$, compared to previous data. The open squares represent the CLAS results for the charged mode with $\Lambda^*$ cuts included \cite{CLAS2014}. The hatched histogram represents systematic errors for this work.}
\end{figure}

The data were divided into three energy bins from 1.67 to 2.27 GeV, six energy bins from 2.27 to 2.87 GeV, and six angular bins from $-0.6$ to $0.0\ \text{GeV}^2$ in $t-t_{\text{min}}$.
Figure~\ref{fig:cs_fit} shows the $t$ dependences of differential cross sections $d\sigma/dt$ in each photon energy bin.
Consistency with the LEPS 2005 results \cite{Mibe2005} in the overlapping energy region is confirmed (Fig.~\ref{fig:cs_fit}), and cross sections for the $\gamma p\rightarrow K^+Y$ ($Y=\Lambda, \Sigma^0$) reactions were also checked to validate the cross section normalization \cite{Sumihama2006}.
Fits to $d\sigma/dt$ distributions were performed using Eq.~(\ref{eq:exp}), with $(d\sigma/dt)_{t=t_{\text{min}}}$ and $B$ as floating parameters (Fig.~\ref{fig:cs_fit}).
The energy dependence of the $t$-slope factor $B$ is shown in Fig.~\ref{fig:bpar}.
The LEPS results show no strong energy dependence of $B$ beyond statistical errors.
The average value of $B$ of this work is $3.57\pm 0.12\ \text{GeV}^{-2}$.
Curves fitted with fixing $B$ at the average value describe the data points well as shown in Fig.~\ref{fig:cs_fit}.
Comparing the combined LEPS results with the CLAS results, the average $B$ value of LEPS results is smaller than that of CLAS results by $21.7 \%$ in the photon energy range of $1.5 < E_{\gamma} < 2.2\ \text{GeV}$ with a statistical significance of $3.2\sigma$.
On the other hand, the LEPS result is larger than the CLAS result by $9.7 \%$ in $2.2 < E_{\gamma} < 2.9\ \text{GeV}$ with $2.4\sigma$.

\begin{figure}[h]
\includegraphics[width=90mm,clip]{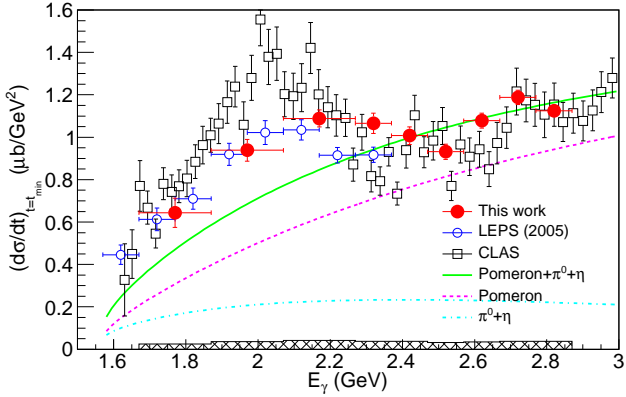}
\caption{\label{fig:cpar} Energy dependence of $(d\sigma/dt)_{t=t_{\text{min}}}$. The red solid circles are the results of the present work. The error bars represent statistical errors. The hatched histogram represents systematic errors. The open squares represent the CLAS results for the charged mode with $\Lambda^*$ cuts included \cite{CLAS2014}. The green solid curve represents the theoretical calculation with the Pomeron strength factor determined by the present measurements. See text for details.}
\end{figure}

The energy dependence of $(d\sigma/dt)_{t=t_{\text{min}}}$ when the $t$-slope factor $B$ is fixed to the average value is shown in Fig.~\ref{fig:cpar}.
Our measurements cover the very forward angle region, therefore, $(d\sigma/dt)_{t=t_{\text{min}}}$ is well determined.
Systematic errors due to the energy dependence of the $t$-slope factor were estimated to be $0.3-2.3\%$.
Comparing with the CLAS results, the LEPS measurements show smaller $(d\sigma/dt)_{t=t_{\text{min}}}$ below $E_{\gamma}=2.2\ \text{GeV}$, and the energy dependence in the nonmonotonic region is more moderate.
The green solid curve represents the theoretical calculations considering $t$-channel exchanges of Pomeron and pseudoscalar $\pi^0$,$\eta$-mesons.
We use the pseudoscalar-meson-exchange amplitudes of Ref.~\cite{Titov2003}.
As for form factors, parameters in Ref.~\cite{Ryu2014} are used.
For the $t$-channel Pomeron exchange process, we use the Donnachie-Landshoff model \cite{Donnachie1984,Donnachie1986,Donnachie1987}.
The invariant amplitude \cite{Titov2003} is given by
\begin{equation}
  I^{\mathbb{P}}_{fi} = -M(s,t)\varepsilon^*_{\mu}(q,\lambda_{\phi})\bar{u}(p',m_f)h^{\mu\nu}_{\mathbb{P}}u(p,m_i)\varepsilon_{\nu}(k,\lambda_{\gamma}),
\label{eq:amp1}
\end{equation}
where $\varepsilon(k,\lambda_{\gamma})$ ($\varepsilon(q,\lambda_{\phi})$) is the polarization vector of the incident photon (outgoing $\phi$ meson) with momentum $k$ ($q$) and spin projection $\lambda_{\gamma}$ ($\lambda_{\phi}$), and $u(p,m_i)$ ($u(p',m_f)$) is the Dirac spinor of the nucleon with momentum $p$ ($p'$) and spin projection $m_i$ ($m_f$).
The vertex function $h_{\mathbb{P}}$ is defined as Eqs. (27,28) of Ref.~\cite{Titov2003}.
The scalar function $M(s,t)$ is described by the following Regge parametrization \cite{Ozaki2009}:
\begin{equation}
  M(s,t) = C_{\mathbb{P}}F_N(t)F_{\phi}(t)\left(\frac{s}{s_{\mathbb{P}}}\right)^{\alpha(t)-1}\exp\left[-\frac{i\pi}{2}\alpha(t)\right],
\label{eq:amp2}
\end{equation}
where $F_N$ and $F_{\phi}$ are the form factors.
We use the form factor parameters of Ref.~\cite{Ryu2014}, with $s_{\mathbb{P}}=4\ \text{GeV}^2$ as in Refs.~\cite{Titov2003,Ryu2014}, and $\alpha(t)=1.08+0.25t$ is the Pomeron trajectory.
Also, $C_{\mathbb{P}}$ is the strength factor.
The previously used strength factors \cite{Titov2003,Ryu2014} were determined by old measurements at higher energies \cite{DESY1978}, which are not consistent with CLAS measurements \cite{CLAS2014} in the overlapping region.
We determined the Pomeron strength factor $C_{\mathbb{P}}$ using our highest-$E_{\gamma}$ data points.
The three highest-$E_{\gamma}$ data points are used, and $C_{\mathbb{P}}=0.649(7)\ \text{GeV}^{-2}$ is obtained by a fit, which is $14\%$ smaller than that of Ref.~\cite{Ryu2014}.
The fitting result does not change more than $1.2\%$ when using between two to seven of the highest-$E_{\gamma}$ data points.
Comparing with theoretical calculations, the data shows a $20-30\%$ excess below $E_{\gamma}=2.27\ \text{GeV}$, suggesting the existence of other processes near threshold.

\begin{figure}[h]
\includegraphics[width=90mm,clip]{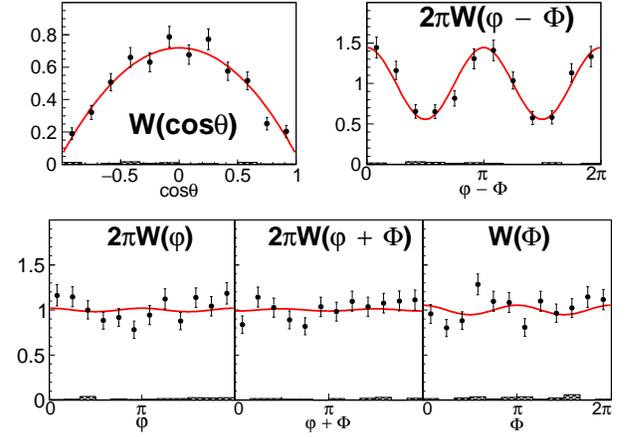}
  \caption{\label{fig:angular} The integrated one-dimensional decay angular distributions in the Gottfried-Jackson frame. The energy and $t$ ranges are $2.37 < E_{\gamma} < 2.77\ \text{GeV}$ and $t-t_{\text{min}}>-0.05\ \text{GeV}^{2}$. The red curves represent the fitting results. The hatched histograms represent systematic errors.}
\end{figure}

\begin{figure}[h]
\includegraphics[width=80mm,clip]{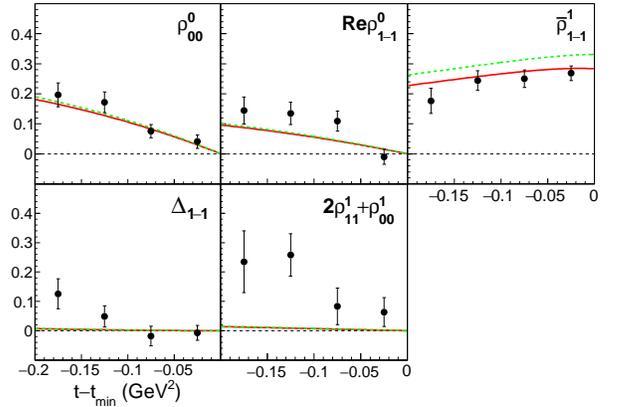}
\caption{\label{fig:rhoVt} $t$ dependences of the spin-density matrix elements in the Gottfried-Jackson frame. The energy range is $2.37<E_{\gamma}<2.77\ \text{GeV}$. The red solid curves represent the theoretical calculations at $E_{\gamma}=2.57\ \text{GeV}$ with the Pomeron strength factor $C_{\mathbb{P}}$ determined by the cross sections ($C_{\mathbb{P}}=0.649\ \text{GeV}^{-2}$). The green dashed curves represent the same model with $C_{\mathbb{P}}=0.7566\ \text{GeV}^{-2}$ \cite{Ryu2014}.}
\end{figure}

The spin-density matrix elements \cite{Schilling1970} were obtained using the following integrated one-dimensional decay distributions:
\begin{eqnarray}
W(\cos\theta) &=& \frac{3}{2}\left(\frac{1}{2}\left( 1-\rho^0_{00}\right)\sin^2\theta +\rho^0_{00}\cos^2\theta\right), \\
W(\varphi) &=& \frac{1}{2\pi}\left(1-2 \text{Re} \rho^0_{1-1}\cos 2\varphi\right), \\
W(\varphi - \Phi) &=& \frac{1}{2\pi}\left(1+2P_{\gamma}\bar{\rho}^1_{1-1}\cos\left[2\left(\varphi - \Phi\right)\right] \right), \\
W(\varphi + \Phi) &=& \frac{1}{2\pi}\left(1+2P_{\gamma}\Delta_{1-1}\cos\left[2\left(\varphi + \Phi\right)\right] \right), \\
W(\Phi) &=& 1 - P_{\gamma}\left(2\rho^1_{11}+\rho^1_{00}\right)\cos 2\Phi.
\end{eqnarray}
Here, $\theta$ and $\varphi$ denote the polar and azimuthal angles of $K^+$ in the Gottfried-Jackson frame (where the spin quantization axis $z$ is parallel to the momentum of the photon in the $\phi$-meson rest frame).
Also, $\Phi$ denotes the angle between the photon-polarization vector and $\phi$-meson production plane, and $P_{\gamma}$ is the degree of polarization of the photon beam, which was derived from the beam energy $E_{\gamma}$ and the degree of polarization of the laser photon.
The validity of $P_{\gamma}$ was confirmed by comparing photon beam asymmetries of hyperon production ($\Lambda$, $\Sigma^0$) with previous results \cite{Sumihama2006}.
The decay angular distributions in the energy and $t$ ranges $2.37<E_{\gamma}<2.77\ \text{GeV}$ and $t-t_{\text{min}}>-0.05\ \text{GeV}^2$ are shown in Fig.~\ref{fig:angular}.
There, $W(\varphi -\Phi)$ shows an oscillation, which indicates the dominance of the natural-parity exchange.
Figure~\ref{fig:rhoVt} shows the $t$ dependences of the spin-density matrix elements in $2.37<E_{\gamma}<2.77\ \text{GeV}$.
The red solid curve represents the theoretical calculations at $E_{\gamma} = 2.57\ \text{GeV}$ using the Pomeron strength factor determined by the cross sections ($C_{\mathbb{P}}=0.649\ \text{GeV}^{-2}$).
The green dashed curve represents the calculations with $C_{\mathbb{P}}=0.7566\ \text{GeV}^{-2}$ \cite{Ryu2014}.
Now $\bar{\rho}^1_{1-1}$ is the most important spin-density matrix element, which is sensitive to the ratio of $t$-channel natural and unnatural parity exchanges, and the theoretical curve using the Pomeron strength factor determined here is closer to the measurements of $\bar{\rho}^{1}_{1-1}$ than the curve using the strength factor in Ref.~\cite{Ryu2014}.
In the large scattering angle region $t-t_{\text{min}}<-0.1\ \text{GeV}^2$, $\Delta_{1-1}$ and the beam asymmetry $2\rho^1_{11}+\rho^1_{00}$ are slightly larger than the theoretical calculations.
In the forward region $t-t_{\text{min}}>-0.1\ \text{GeV}^2$, the theoretical model (Pomeron$+\pi^0+\eta$) well reproduces the measured spin-density matrix elements.

\begin{figure}[h]
\includegraphics[width=80mm,clip]{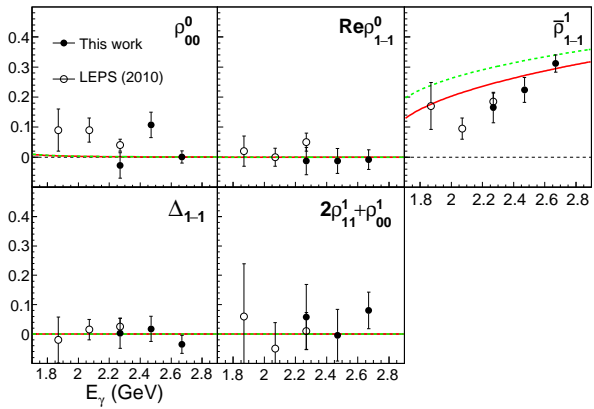}
\caption{\label{fig:rhoVe} $E_{\gamma}$ dependences of the spin-density matrix elements in the Gottfried-Jackson frame. The $t$ range is $t-t_{\text{min}}>-0.05\ \text{GeV}^2$. The open circles represent the previous LEPS results \cite{Chang2010}. The red solid curves represent the theoretical calculations at zero degrees ($t=t_{\text{min}}$) with the Pomeron strength factor $C_{\mathbb{P}}$ determined by the cross sections ($C_{\mathbb{P}}=0.649\ \text{GeV}^{-2}$). The green dashed curves represent the same model with $C_{\mathbb{P}}=0.7566\ \text{GeV}^{-2}$ \cite{Ryu2014}.}
\end{figure}

Figure~\ref{fig:rhoVe} shows $E_{\gamma}$ dependences of the spin-density matrix elements in the forward region $t-t_{\text{min}} > -0.05\ \text{GeV}^2$.
The results are consistent with previous results of LEPS \cite{Chang2010} in the overlapping energy region.
Note that $\text{Re}\rho^0_{1-1}$ and the photon beam asymmetry $2\rho^1_{11}+\rho^1_{00}$ must go to zero at zero degrees ($t=t_{\text{min}}$) by definition, and the measured values are consistent with zero within the statistical uncertainty.
As for $\bar{\rho}^1_{1-1}$, the data points in the high energy region $E_{\gamma} > 2.37\ \text{GeV}$ are well described by the Pomeron and pseudoscalar exchange model, and the data point in $1.97 < E_{\gamma} < 2.17\ \text{GeV}$ significantly deviates from the model prediction with a statistical significance of $3.4\sigma$.
This fact suggests that additional amplitudes or interferences between the Pomeron exchange and other processes appear near threshold.

In summary, the cross sections and decay asymmetries for the $\gamma p\rightarrow\phi p$ reaction have been measured at SPring-8/LEPS in the photon energy range of $1.5 - 2.9$ GeV.
The $t$-slope factor $B$ does not show a strong energy dependence beyond statistical errors.
We determined the strength factor of the Pomeron exchange using the measured $(d\sigma/dt)_{t=t_{\text{min}}}$ at $E_{\gamma}>2.57\ \text{GeV}$.
Both $(d\sigma/dt)_{t=t_{\text{min}}}$ and the decay asymmetries in the higher energy region ($E_{\gamma}>2.37\ \text{GeV}$) are well reproduced by the Pomeron and pseudoscalar exchange model using the Pomeron strength factor determined here.
In the lower $E_{\gamma}$ region ($E_{\gamma}<2.37\ \text{GeV}$), an excess of $(d\sigma/dt)_{t=t_{\text{min}}}$ is seen compared with the model prediction of $t$-channel exchanges of the standard Pomeron, $\pi^0$ and $\eta$.
In this energy region, the measured spin-density matrix elements $\bar{\rho}^1_{1-1}$ are also not consistent with the model prediction.
These facts suggest the existence of additional processes such as rescattering or additional gluonic processes.
The predominantly imaginary Pomeron-exchange amplitude (see Eqs.~(\ref{eq:amp1}) and (\ref{eq:amp2})) at lower energies is not trivial because of our lack of knowledge of the Pomeron in the low energy region, and it is also possible that the Pomeron-exchange amplitude interferes with other amplitudes near threshold.
To pin down the natural-parity Pomeron-exchange amplitude, a measurement of coherent production from ${}^4\text{He}$ would be useful \cite{Hiraiwa2017}.
Also, $\phi$ photoproduction from deuterons is helpful to understand the production mechanism.
For example, coherent production can be used to extract $\eta$ and Pomeron exchange contributions \cite{Chang2008}, and the ratio of the production rate of neutrons to that of protons in incoherent production can be used to disentangle the $\pi^0$, $\eta$ and Pomeron exchange amplitudes. \cite{Titov2007,Chang2010,ChangMiyabe2010}.
Precise measurements of these reactions and an understanding of the Pomeron-exchange amplitude at lower energies are desired.

We thank the SPring-8 staff for providing excellent experimental conditions.
We thank A. Hosaka, H.-Y. Ryu, and H. Nagahiro for fruitful discussions.
This work is supported by a Grant-in-Aid for JSPS Fellows (No.~251348) and JSPS KAKENHI Grant Number 16H06007.


\providecommand{\noopsort}[1]{}\providecommand{\singleletter}[1]{#1}%

\end{document}